\documentclass[
    ,final            
  ]
  {aipproc}

\layoutstyle{6x9}
\usepackage{bm}

\newcommand{\ba}{\begin{eqnarray*}}
\newcommand{\ea}{\end{eqnarray*}}

\newcommand{\be}{\begin{equation}}
\newcommand{\ee}{\end{equation}}

\newcommand{\beq}{\[} 
\newcommand{\eeq}{\]}

\newcommand{\bsigma}{\mbox{\boldmath $\sigma$}}

\newcommand{\bk}{{\bf k}}

\newcommand{\bp}{{\bf p}}
\newcommand{\bq}{{\bf q}}

\newcommand{\bs}{{\bf s}}

\newcommand{\bP}{\mbox{\boldmath $P$}} 
\newcommand{\bQ}{\mbox{\boldmath $Q$}} 
 
\newcommand{\bfsigma}{\mbox{\boldmath $\sigma$}}

\def\lsim{\mathrel{\rlap{\lower4pt\hbox{\hskip1pt$\sim$}}
    \raise1pt\hbox{$<$}}}         
\def\gsim{\mathrel{\rlap{\lower4pt\hbox{\hskip1pt$\sim$}}
    \raise1pt\hbox{$>$}}}         
\def\beq{\begin{equation}}
\def\eeq{\end{equation}}
\def\bea{\begin{eqnarray}}
\def\eea{\end{eqnarray}}

\newcommand{\text}{\hbox}

\newcommand{\open}{{<\kern -0.3 em{\scriptscriptstyle )}}}

\begin{document}

\title{\mbox{}\\[-1.8 cm] Spin Filtering of Stored (Anti)Protons:\\
 from FILTEX to COSY to AD to FAIR 
\footnote{Presented at the 17th Intl. Spin Physics Symposium,
SPIN2006, Kyoto, Japan, October 2-7, 2006}}

\classification{24.70.+s, 29.27.Hj\\[-6 mm]}

\author{Nikolai Nikolaev and Fyodor Pavlov\\[-2 mm]}{address={
Institut f\"ur Kernphysik, Forschungszentrum J\"ulich, 52428 J\"ulich, Germany
}
}

\begin{abstract}
We review the theory of spin filtering of stored 
(anti)protons by multiple
passage through the polarized internal target (PIT). 
Implications for the antiproton 
polarization buildup in the proposed PAX experiment at FAIR
GSI are discussed. 
\end{abstract}

\maketitle


 An ambitious physics program with polarized 
antiproton--polarized proton collider has been proposed recently by 
the PAX Collaboration
\cite{PAXTP} for 
FAIR at GSI in Darmstadt, Germany \cite{Hans}. Such a collider would give an
unique access to the last
leading--twist missing piece of the QCD  partonic
structure of the nucleon  --- the transversity --- 
which can only be investigated via
double--polarized $\bar{p}p$ Drell--Yan production and without which
the spin tomography of the proton
would be ever incomplete. At the core of the PAX proposal is spin filtering of stored
antiprotons by multiple passage through a Polarized Internal hydrogen
gas Target  (PIT) \cite{PAXTP,PAXPRL} --- a technique tested
by the FILTEX experiment at 23 MeV proton 
TSR-ring in Heidelberg \cite{FILTEX}. In its extension  
to antiprotons there remain open issues, though.

In his theory of the FILTEX result,
H.~O. Meyer (i)  obsreved that stored particles which scatter
elastically in PIT at angles within a storage ring acceptance angle 
 $\theta_{\mathrm{acc}}$ are retained in the beam and their
polarization complements the polarization by transmission 
and (ii) argued that the QED 
polarization transfer from polarized target electrons
to scattered protons \cite{HM}
is crucial for the qunatitative understanding of the FILTEX 
result \cite{Meyer}. This  prompted an idea to base the antiproton
polarizer of the PAX on spin filtering by 
polarized electrons in PIT \cite{PAXPRL}.

After the PAX proposal, the interplay of the transmission and 
Scattering Within the Ring Acceptance Angle (SWRAA) mechanisms, 
and the feasibility of  filtering on electrons,
became a major issue. Yu. Shatunov
\footnote{Yu. Shatunov, private communication} was
perhaps the first to question the filtering by electrons. 
Eventually two groups  of theorists --- at the Budker Institute
\cite{MS}
and IKP, J\"ulich 
\cite{KolyaFedya} --- came to a conclusion on the self-cancellation of the 
polarized electron contribution 
to the spin filtering of (anti)protons. Here we present a brief review
of this finding and its implications for the PAX program.

There is an important hierarchy of scattering angles 
$\theta$ in the proton-atom scattering.
First, the Coulomb fields of the proton and atomic electron screen 
each other for scattering angles $
\theta < \theta_{\mathrm{min}} =
{ \alpha_{em} m_e / \sqrt{2m_p T_p}} \approx 2\cdot 10^{-2}\quad {\rm mrad}$
 ($T_p=23$ MeV).
Second, light electrons do not deflect protons, $
\theta \leq { \theta_e ={m_e/  m_p}} \approx 5\cdot 10^{-1}\quad {\rm mrad}$.
Third comes the Colomb-Nuclear Interference (CNI) angle
$\theta_{CNI} 
\approx \sqrt{{2\pi \alpha_{em}/m_p T_p \sigma_{tot,nucl}^{pp}}}
\approx {100 {\rm mrad}}$. Fourth 
for the TSR $\theta_{acc}=4.4 \quad {\rm mrad}$, and $
\theta_{min} \ll { \theta_e} \ll { \theta_{acc}} \ll \theta_{CNI}$.
 For important 
angles $\theta > \theta_{\mathrm{min}}$ the prroton-atom interaction is
dominated by quasielastic (QE) scattering, ${  p}+
{{ } atom} \to{  p'_{scatt}} +e_{spect}+p_{recoil}, 
{  p'_{scatt}} +p_{spect}+e_{recoil}$ ($\bq$ is the 
momentum transfer, $\hat{\rho}$ --- the beam spin-density matrix)
):
$$
{d\hat{\sigma}_{QE} \over d^2\bq} =
{1\over (4\pi)^2} \hat{\cal F}(\bq){{ }\hat{\rho}}\hat{\cal F}^{\dagger}(\bq)=
{1\over (4\pi)^2} \hat{\cal F}_{
 e}(\bq){{ }\hat{\rho}}\hat{\cal F}_{
 e}^{\dagger}(\bq)
+{1\over (4\pi)^2} \hat{\cal F}_{
 p}(\bq){{ }\hat{\rho}}\hat{\cal F}_{
 p}^{\dagger}(\bq)
$$
In our normalization the forward scattering amplitude 
$\hat{F}(0) = \hat{R}(0) + i\hat{\sigma}_{tot}$. For 
spin-${1\over 2}$ beam and target $
 \hat{\sigma}_{tot}= \sigma_0 +\sigma _1({
\bfsigma} \cdot {{ }\bQ})+
 \sigma_2 ({
\bfsigma}\cdot \bk)( {{ }\bQ}\cdot \bk)$, where 
 $\bk$ is the beam axis, $\bQ$ and $\bP$ are the target 
and beam polarizations.

Let $N$ be the volume density of atoms in PIT and $z$ the integrated thickness
of the PIT for a circulating particle. The spin-momentum density matrix of
the beam,  
$
\hat{\rho}(\bp)={1\over 2}[I_0(\bp) + {\bsigma} {
\bs(\bp)}],
$
satisfies the evolution equation 
\begin{eqnarray}
{d\over dz}\hat\rho &
= &\underbrace{i{1\over
2}N\Big(\hat{R} \hat{\rho}(\bp) -
\hat{\rho}(\bp)\hat{R}\Big)}_{\rm Precession~in~transmission}
- 
\underbrace{{1\over 2}N
\Big(\hat{\sigma}_{tot}\hat{\rho}(\bp)+\hat{\rho}(\bp)
\hat{\sigma}_{tot}\Big)}_{
 {\rm Filtering~by ~ transmission}} \nonumber\\
&+&  \underbrace{N\int^{\Omega_{acc}} {d^2\bq \over (4\pi)^2}
\hat{\cal F}(\bq)\hat{\rho}(\bp-\bq) \hat{\cal F}^{\dagger}(\bq)}_
{
{\rm Feedback~ from~ SWRAA}}\label{eq:EvolEqn}
\end{eqnarray}
The stable polarization are either normal to ring plane (the case in the
FILTEX experiment)  
or longitudinal if a
ring is furnished with the Siberian Snakes. The precession effects are very
important in the polarized neutron optics but average out in our case. 
Upon 
neglecting the precession
terms $\propto \hat{R}$,  Eq. (\ref{eq:EvolEqn})  
boils down to the 
kinetic equation for spin population numbers. For real storage rings Eq.
(\ref{eq:EvolEqn})
further simplifies because the angular divergence of the beam at PIT is 
much smaller than $\theta_{acc}$.
The FILTEX PIT used the hyperfine state with
parallel proton and electron polarizations. 

The real issue is a pattern of a (partial) cancellation of transmission 
and SWRAA effects in Eq. (\ref{eq:EvolEqn}). 
Without spin-flip, the polarization buildup follows 
$
 P(z)= -\tanh( Q \sigma_P Nz)
$ where $\sigma_P =\sigma_1$. Because only those particles 
which scatter in PIT 
at angles $\theta >\theta_{\mathrm{acc}}$ are removed from the stored 
beam,  Meyer argued that the transmission be evaluated taking 
$
\hat{\sigma}_{tot} = \hat{\sigma}_{tot}(\theta_{\mathrm{acc}} <\theta)$.
 For all-angle nuclear interaction 
without CNI, the SAID 
phase shifts give  $\sigma_{1,nuclear} =122$ mb, upon the 
correction for CNI Meyer found 
$\sigma_1({ CNI;}\theta >\theta_{\mathrm{acc}})= 83$ mb
vs. the published FILTEX result $\sigma_P(FILTEX, 1993) = 63 \pm 3$ (stat.) mb.
Next Meyer includes the polarization from SWRAA. In view of 
$\theta_e \ll \theta_{acc}$, the scattering off electrons is entirely 
 SWRAA and contributes  $\delta\sigma_1^{ep}(\theta <
\theta_{acc})= -70$mb. SWRAA off protons 
contributes  $\delta\sigma_1^{pp}(\theta <
\theta_{acc}) = +52$ mb. Meyer's net result 
for the polarization cross section \cite{Meyer}, 
\beq
\sigma_P =  \sigma_1({ CNI;} \theta >\theta_{\mathrm{acc}})+
  \delta\sigma_1^{pp} +  \delta\sigma_1^{ep}
= 135 \quad{\rm mb} +   \delta\sigma_1^{ep} = 65\quad {\rm mb}, 
\label{MeyerSigmaP}
\eeq
is in perfect agreement with the published FILTEX result. 
A subsequent reanalysis of the target density and polarization gave 
$\sigma_P(FILTEX, 2004) = 72.5 \pm 5.8
 (stat. + sys.)$ (F.Rathmann, see \cite{PAXTP}).

The Budker and J\"ulich groups argue that $\hat{\sigma}_{tot}$
in the transmission term must rather include a scattering on 
atoms at all angles $\theta > \theta_{\mathrm{min}}$: 
$
\hat{\sigma}_{tot} = \hat{\sigma}_{tot}(\theta_{\mathrm{min}} >\theta)=
\hat{\sigma}_{tot}(\theta_{\mathrm{min}} <\theta < \theta_{\mathrm{acc}})+
\hat{\sigma}_{tot}(\theta_{\mathrm{acc}} <\theta)$.
Then one would readily find that the beam polarization-independent 
SWRAA cancels exactly 
the corresponding transmission effects from 
$\hat{\sigma}_{tot}(\theta_{\mathrm{min}} <\theta < \theta_{\mathrm{acc}})$.
For a polarized beam  there is a
mismatch between the spin-filtering component in 
$\hat{\sigma}_{tot}(\theta_{\mathrm{min}} <\theta < \theta_{\mathrm{acc}})$
and the polarization feedback from SWRAA. This mismatch is entirely due 
the spin-flip elastic proton-atom scattering at angles 
$\theta < \theta_{\mathrm{acc}}$.
The generic solution for the polarization buildup reads
\begin{equation}
P(z)= -{Q(\sigma_1 +{
 \Delta\sigma_1})\tanh(Q\sigma_3 Nz )
\over Q\sigma_3 +{0.5 \Delta\sigma_0} \tanh(Q\sigma_3Nz)},
\label{eq:BudkerJuelich}
\end{equation}
where  $\Delta\sigma_{0,1}$ are the proton spin-flip (SF)
cross sections for 
an unpolarized
and polarized target, respectively, $ |\Delta\sigma_{1}|\leq
|\Delta\sigma_{0}|$, and  ${ Q}\sigma_3 = \sqrt{{{ } Q}^2\sigma_1(\sigma_1+{
 \Delta\sigma_1})+{   \Delta\sigma_0}^2/ 4}$.
The formulas for  $\Delta\sigma_{0,1}$ in terms of the two-spin observables
are found in \cite{MS,KolyaFedya}. In 
contrast to the Meyer approach, in the
Budker-J\"ulich  analysis the 
electron-to-proton polarization transfer is entirely canceled by the electron
contribution to the transmission filtering. 
For nuclear SF scattering at $\theta \leq \theta_{acc} \ll \theta_{CNI}$ CNI
is arguable negligible and a crude estimate is
$\Delta\sigma_0 \lsim \sigma_{tot} \theta_{\mathrm{acc}}^2
\lsim 10^-4 \sigma_{tot}$. Within the Budker-J\"ulich approach, the 
small-time polarization buildup is controlled by (SAID-SP05 database)
\begin{equation}
\sigma_{P}\approx  - (\sigma_1( CNI; \theta>\theta_{\mathrm{acc}}) +
 \Delta\sigma_1)=85.6~{\rm mb}.
\label{SigmaBudkerJuelich}
\end{equation}

The case of the pure electron target deserves a special consideration. 
Here $
\sigma_0(\theta>\theta_{\mathrm{acc}})=0,\quad 
\sigma_1(\theta>\theta_{\mathrm{acc}})=0,\quad
Q\sigma_3 = {  \Delta\sigma_0}$, the beam is not attenuated and the
polarization
buildup follows 
\beq
{ P(z)} = { P(0)} \exp(-N{ \Delta\sigma_0 }z) +
{ Q}{{ \Delta\sigma_1} \over { \Delta \sigma_0}}
\Big\{1-\exp(-N{ \Delta\sigma_0}z)\Big\}.
\label{eq:ElectronTarget}
\eeq
Spin filtering by electron coolers was
discussed in the PAX TP 
with the conclusion that the attainable target densities 
are too low \cite{PAXTP}. More recently, Th. Walcher et al. 
\footnote{Th.Walcher, private
communication} argued that if the SF on electrons
is comparable to the electron-to-proton
spin transfer, then filtering in a pure electron target can be enhanced
considerably 
by a judicious  choice of the non-relativistic relative velocity of
the comoving electron and proton beams. At the moment, based only on the 
FILTEX result, one can not discriminate between the Meyer and 
Budker-J\"ulich treatments of the electron contribution to filtering;
as it is a common practice with conflicting theories, the issue 
must be clarified experimentally.

First, the filtering by SF has never been tested experimentally. 
If the polarized electron target polarizes
the initially unpolarized stored beam, the unpolarized electron target 
depolarizes the stored proton beam, see Eq. \ref{eq:ElectronTarget}. 
The required density of electrons 
is provided by the  $^4$He internal target, which has an advantage of the
spin-0 nucleus. Consequently, a useful upper bound on depolarization 
by electrons, i.e., $\Delta\sigma_0$ and
$| \Delta\sigma_1| \leq |\Delta \sigma_0|$ can be deduced. 
Such an experiment, the idea of which grew up from discussions
with H.O. Meyer, is being planned at COSY \cite{COSYSF}. Second, 
filtering on electrons and on protons have a very distinct 
energy dependence. In Fig. 1 we show the predictions from the
Budker-J\"ulich approach for the nuclear
spin filtering cross section which can be tested at COSY.
The confirmation of this energy dependence 
would be a convincing proof that spin filtering is dominated by 
nuclear interaction of a negligible filtering on electrons.

\begin{figure}[!t]
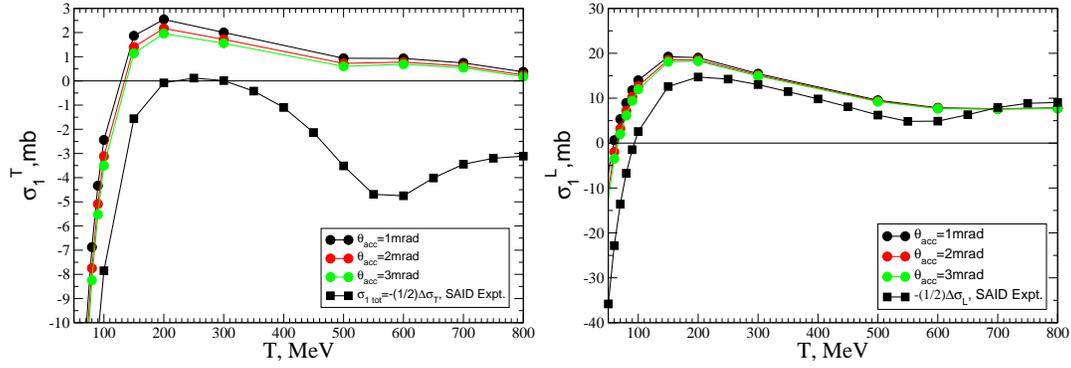

   \centering
\includegraphics[width = 7.0cm]{SG1TAll.eps}
\includegraphics[width = 7.0cm]{SG1LAll.eps}
\caption{Predictions for spin filtering of protons by nuclear 
interaction in polarized hydrogen PIT for transverse (T) and longitudinal
(L) polartizations. The curves marked by back squares are for pure
nuclear $pp$ interaction, the 
three curves for different range acceptance angle are predictions 
with allowance for CNI.}
\end{figure}

We come to a summary. FILTEX experiment is  an
important  proof of the principle of spin filtering. The Meyer and
 Budker-J\"ulich approaches disagree in the treatment of SWRAA and
significance of
the electron contribution to spin filtering. If the electrons do not
contribute (Budker-J\"ulich), then filtering of antiprotons would
depend on spin-dependence of $\bar{p}p, \bar{p}D$ interactions.
The existing models of $\bar{N}N$ interactions are encouraging but 
not reliable because of a lack of double-spin observables to fix the
model parameters. The solution for PAX is to optimize the filtering
energy with antiprotons available at existing facilities (CERN AD) \cite{COSYSF}.
The experimental constraints on the electron contribution 
to filtering can be obtained from proton depolarization and 
energy-dependence of filtering of protons at COSY.



\end{document}